\documentclass[prb,preprint,showpacs,eqsecnum]{revtex4}
\usepackage{graphicx}
\begin{document}
\title{Magnetic relaxation in finite two-dimensional
nanoparticle ensembles}
\author{S. I. Denisov}
\email{denisov@ssu.sumy.ua}
\author{T. V. Lyutyy}
\affiliation{Department of Mechanics and Mathematics,
Sumy State University, 2, Rimskiy-Korsakov Street, 40007 Sumy,
Ukraine}
\author{K. N. Trohidou}
\email{trohidou@ims.demokritos.gr} \affiliation{Institute of
Materials Science, NCSR ``Demokritos,'' 15310 Athens, Greece}

\begin{abstract}
We study the slow phase of thermally activated magnetic relaxation
in finite two-dimensional ensembles of dipolar interacting
ferromagnetic nanoparticles whose easy axes of magnetization are
perpendicular to the distribution plane. We develop a new method
to numerically simulate the magnetic relaxation for the case that
the smallest heights of the potential barriers between the
equilibrium directions of the nanoparticle magnetic moments are
much larger than the thermal energy. Within this framework, we
analyze in detail the role that the correlations of the
nanoparticle magnetic moments and the finite size of the
nanoparticle ensemble play in magnetic relaxation.

\end{abstract}

\maketitle

\section{Introduction}

The role of the dipolar interaction in systems of nanometer-sized
ferromagnetic particles, or nanoparticle ensembles for short, has
been intensively studied in recent years. Such ensembles have
numerous technological applications, and it is important to
understand their magnetic phenomena and
processes.\cite{DFT97,HS94} One of the most complicated problems,
where dipolar interactions must be taken into account, is
thermally activated magnetic relaxation. To derive the law of
magnetic relaxation, i.e., of the dimensionless reduced
magnetization, usually requires the derivation of the distribution
function of the nanoparticle magnetic moments. In the simplest
case, that of non-interacting nanoparticles with conserved total
magnetic moments, the distribution function obeys the
Fokker-Planck equation.\cite{B63} For nanoparticle ensembles with
more or less realistic magnetic energy, however, its
time-dependent solutions are not known, and exact results were
found mainly for numerical characteristic of the relaxation
process such as the largest relaxation time.
\cite{B63,KG90,B93,G96,C_etc98,K00}

Unfortunately, in the case of dipolar interacting nano\-particles
no exact results for the magnetic relaxation exist. This fact
makes it difficult to check the validity of different approximate
methods and approaches that are extensively used in this
area.\cite{SW81,DBF88,MT94,HM98,DFT99,PYPG99,D99,ESA00,Z_etc01}
The justification of approximations is a very important task
because the use of non-rigorous, although plausible,
approximations can lead to opposite conclusions.\cite{HM98,DFT99}
One expects that a sufficiently rigorous analysis of the
relaxation law can be performed for the simplest systems like
two-dimensional (2D) ensembles of identical, spherical
nanoparticles with conserved magnetic moments and large uniaxial
perpendicular anisotropy. Such ensembles represent an important
class of perpendicular magnetic recording media,\cite{PEW01} and
they are convenient systems to study experimentally and
theoretically the role that the dipolar interaction plays in
magnetic relaxation.

Magnetic relaxation in such ensembles was considered first by
Lottis, White, and Dahlberg\cite{LWD91} within the simplified
version of the mean-field approximation. Using the concept of a
demagnetizing field, the authors wrote down the equation that
describes the relaxation of magnetization from the initial state,
when all nanoparticle magnetic moments are oriented along a
certain direction of the easy axis, to the demagnetized ground
state. They solved this equation numerically and showed that for a
limited time domain relaxation occurs slower than the Debye model
predicts. They approximated the relaxation law $\rho(t)$ by a
stretched-exponential dependence, which, however, does not hold
for all times.

Recently, we studied the influence of the mean and the fluctuating
components of the dipolar field on the process of magnetic
relaxation in those ensembles.\cite{DT01,DT02} Using the
Fokker-Planck equation, we derived an equation that describes the
so-called slow relaxation, i.e., relaxation for times exceeding
the time $t_{qe}$ to establish the quasiequilibrium distribution
of the magnetic moments ($t_{qe}\sim10^ {-8}$~s, see Sec.~II B),
and we solved it in limiting cases. We showed that both the mean
and the fluctuating components of the dipolar field enhance
relaxation, and that for small and large times magnetic relaxation
has a Debye character, but the corresponding relaxation times can
be very different. This difference causes the quasi-logarithmic
relaxation at intermediate times that was found numerically  in
Ref.~\onlinecite{LWD91}.

The role that the correlations of directions of the nanoparticle
magnetic moments play in magnetic relaxation has not yet been
clarified. Clearly, correlation effects are very significant, and
we expect that, due to the antiferromagnetic character of the
dipolar interaction in such ensembles, they can qualitatively
change the relaxation law. The influence of the finite size of the
nanoparticle ensemble on magnetic relaxation is another important
problem, which also has not yet been addressed. We expect that,
due to the long-range character of the dipolar interaction,
magnetic relaxation will significantly depend on the ensemble
size, especially for small times when dipolar fields near the
internal and the external magnetic moments can be quite different.

The complexity of these problems forces us to seek numerical
solutions. The known methods of numerical simulation of magnetic
relaxation, such as directly integrating
the stochastic Landau-Lifshitz equation,\cite{LC93,GL98,%
BGG02} the conventional Monte Carlo method,\cite{GPRB00,%
S_etc01} and the time quantified Monte Carlo method\cite{NCK00}
are not suitable for our purposes. The main reasons are the
following. To integrate the Landau-Lifshitz equation, the
integration time step must be smaller than the inverse of the
precession frequency of the nanoparticle magnetic moments
($\sim10^ {-11}$~s). Therefore, this method usually works only for
the description of the fast magnetic relaxation, i.e., relaxation
on time scales smaller than $t_{qe}$. The conventional Monte Carlo
method is not suitable, since each Monte Carlo step has no
physical time associated with it. The time quantified Monte Carlo
method also cannot be applied to our situation; the number of
Monte Carlo steps that are necessary to calculate the relaxation
law on times comparable with the relaxation time becomes
prohibitively large in the case of high potential barriers between
the equilibrium directions of the magnetic moments (see Sec.~III
B). Further, that method is valid only in the high damping limit,
i.e., if there is no precession of the magnetic moments.

In this paper we develop a new method to numerically simulate
thermally activated magnetic relaxation in finite 2D ensembles of
dipolar interacting ferromagnetic nanoparticles. We consider the
case where the nanoparticles with uniaxial anisotropy occupy the
sites of a square lattice and their easy axes of magnetization are
perpendicular to the lattice plane. We develop an equation that
relates the magnetization of the ensemble at the next time step to
the known state of the nanoparticle ensemble at the previous time
step, and a numerical procedure that defines the ensemble state at
the next time step. To derive the probability densities for the
reorientation of the nanoparticle magnetic moments, contained in
this equation, we exploit that they can be represented via the
mean times for magnetic moments to reorient, or, in other words,
via the so-called mean first-passage times, and calculate these
times using the backward Fokker-Planck equation.

The paper is organized as follows. In Sec.~II, we introduce the
equation mentioned above  and derive rigorous expressions for the
probability densities of reorientation of the nanoparticle
magnetic moments. The algorithm for the numerical calculation of
the relaxation law is described in Sec.~III. In the same section
we present the numerical results and analyze the features of the
magnetic relaxation caused by both the correlations of the
nanoparticle magnetic moments and the finiteness of the
nanoparticle ensemble. We summarize our results in Sec.~IV.

\section{Analytical results}

We consider a system of $N$ uniaxial and identical spherical
ferromagnetic nanoparticles with a radius $r$. We assume that the
nanoparticle centers occupy the sites of a square lattice of size
$Ld\times Ld$ [$(L+1)^{2}=N$] and lattice spacing $d(\geq 2r)$.
The easy axes of nanoparticles magnetization are perpendicular to
the lattice plane ($xy$-plane), and at the initial time $t=0$ all
magnetic moments $\textbf{m}_{i}(t)$ (the index $i$ labels the
nanoparticles) are oriented along the $z$-axis (see Fig.~1). We
also assume that the smallest heights $\Delta U_{i}$ of the
potential barriers between the equilibrium directions of the
nanoparticle magnetic moments are much larger than the thermal
energy $k_{B}T$ ($k_{B}$ is the Boltzmann constant, $T$ is the
absolute temperature), i.e., the condition $\varepsilon_{i}=\Delta
U_{i}/k_{B}T\gg1$ holds for all nanoparticles. The main goal of
this section is to find the relation between the reduced
magnetization at times $t$ and $t+\tau$.

\subsection{Equation for the reduced magnetization}

For $\varepsilon_{i}\gg1$, the vectors $\textbf{m}_{i}(t)$
fluctuate within small vicinities of the positive and negative
directions of the $z$-axis, and they are reoriented only rarely.
Consequently, the average numbers of positively and negatively
oriented magnetic moments have well-defined values $N_{+}(t)$ and
$N_{-}(t)$, respectively, at any instant $t$. Since the number of
magnetic moments that at time $t$ have reoriented is much less
than $N$, the approximate relation $N_{+}(t)+N_{-}(t)\approx N$
holds, and we can define the reduced magnetization of the
nanoparticle ensemble as $\rho(t)=2N_{+}(t)/N-1$. Let us define
also the state of that ensemble. We assume that the state of the
nanoparticle ensemble at time $t$ is known if the directions of
all magnetic moments are known, i.e., we describe the ensemble
state by the set of signs $\sigma_{i}(t)\equiv\sigma_{i}$
($i=1,...,N$), where $\sigma_{i}=+$ or $-$ depending on whether
the vector $\textbf{m}_{i}(t)$ fluctuates around the positive or
negative direction of the $z$-axis.

Given the ensemble state, neglecting the fluctuations of
$\textbf{m}_{i}(t)$, and taking into account that approximately
$\textbf{m}_{i}(t)= \sigma_{i}m\textbf{e}_{z}$ for the time
intervals between the reorientations, we can write the local
dipolar field $\textbf{h}_{i}(t)$ acting on the magnetic moment
$\textbf{m}_{i}(t)$ as $\textbf{h}_{i}(t)=
h_{i}(t)\textbf{e}_{z}$. Here
\begin{equation}
       h_{i}(t)=-m\sum_{j\neq i}\sigma_{j}\frac{1}{r_{ij}^{3}}\,,
       \label{eq:dipfield}
\end{equation}
$m=|\textbf{m}_{i}(t)|$, $\textbf{e}_{z}$ is the unit vector along
the $z$-axis, and $r_{ij}$ is the distance between the centers of
corresponding nanoparticles. If at time $t$ the magnetic moments
do not undergo reorientations, then each nanoparticle is under the
influence of the local dipolar field (\ref{eq:dipfield}). Even if
some magnetic moments are reoriented, their number is much less
than $N$ because $\varepsilon_{i}\gg1$, and formula
(\ref{eq:dipfield}) remains approximately valid. For sufficiently
small times intervals we can consider therefore the ensemble of
interacting nanoparticles as a system of independent magnetic
moments, each of which feels its own external magnetic field
$\textbf{h}_{i}(t)$. This fact significantly simplifies the
numerical investigation of the magnetic relaxation in ensembles of
dipolar interacting nanoparticles.

Let us assume that the probabilities of reorientation per unit
time $w_{\sigma_{j}}(t;j)$ (i.e., the probability densities of
reorientation) of the vectors $\textbf{m}_{j}(t)$ ($j=1,...,N$)
from the positive direction of the $z$-axis (if $\sigma_{j}=+$)
and from the negative one (if $\sigma_{j}=-$) are known. We also
assume that on the interval $(t,t+\tau)$ the probabilities of two
and more reorientations of $\textbf{m}_{j}(t)$ are negligibly
small. Then, taking into account that $N_{+}(t+\tau) - N_{+}(t)$
is equal to the difference between the number of reorientations
from the negative direction of the $z$-axis and the number of
reorientations from the positive direction of the $z$-axis, we
obtain
\begin{equation}
       \rho(t+\tau)-\rho(t)=-\frac{2\tau}{N}\sum_{j=1}^{N}
       \sigma_{j}w_{\sigma_{j}}(t;j)\,.
       \label{eq:difference}
\end{equation}

The probability densities $w_{\sigma_{j}}(t;j)$ depend on the
local field $h_{j}(t)$, and Eq.~(\ref{eq:difference}) can be
applied if the ensemble state at time $t$ is known. However
Eq.~(\ref{eq:difference}) is not an iterative equation for the
ensemble state; it only defines $\rho(t+\tau)$ but not the
ensemble state at time $t+\tau$. In order to use
Eq.~(\ref{eq:difference}) as the recurrence equation for the
calculation of the law of magnetic relaxation, we need to
determine the values $w_{\sigma_{j}}(t;j)$ and develop a procedure
to find the state of the nanoparticle ensemble at time $t+\tau$,
if its state at time $t$ is known. We will describe that procedure
in the next section. Below we calculate the probability densities
$w_{\sigma_{j}}(t;j)$.

\subsection{Probability densities of reorientation}

The probability densities of reorientation are given by
$w_{\sigma_{j}}(t;j)= 1/t_{s}^{\sigma_{j}}(t;j)$, where
$t_{s}^{\sigma_{j}}(t;j)$ are the mean times that the magnetic
moment $\textbf{m}_{j}(t)$ spends pointing in the positive (when
$\sigma_{j}=+$) and the negative (when $\sigma_{j}=-$) directions
of the $z$-axis. These times can be represented as
$t_{s}^{\sigma_{j}}(t;j) = 2t_{m}^{\sigma_{j}}(t;j)$, where
$t_{m}^{\sigma_{j}}(t;j)$ are the mean times for
$\textbf{m}_{j}(t)$ to reach for the first time the state with a
maximum value of the nanoparticle magnetic energy $W_{j}$. The
factor 2 takes into account the fact that from that state the
magnetic moment $\textbf{m}_{j}(t)$ can transit to the state
$\sigma_{j}=+$ or $\sigma_{j}=-$ with probability $1/2$. In our
case, the magnetic energy $W_{j}$ includes the anisotropy energy
$-(H_{a}/2m) m_{jz}^{2}(t)$ and the Zeeman energy $-h_{j}(t)
m_{jz}(t)$, so that it has axial symmetry and
\begin{eqnarray}
       W_{j}&\equiv& W_{j}(\theta_{j}(t),t)
       \nonumber\\
       &=&-\frac{1}{2}H_{a}m\left[\cos^{2}\theta_{j}(t)+
       2b_{j}(t)\cos\theta_{j}(t)\right].
       \label{eq:energy}
\end{eqnarray}
Here $H_{a}$ is the anisotropy field, $\theta_{j}(t)$ is the polar
angle of $\textbf{m}_{j}(t)$, and $b_{j}(t)=h_{j}(t)/H_{a}$
(assuming that two equilibrium directions exist for each magnetic
moment, $|b_{j}(t)|<1$ for all nanoparticles). Accordingly, the
state corresponding to the maximum value of $W_{j}$ is defined by
the polar angle
\begin{equation}
       \Omega_{j}(t)=\arccos\left[-b_{j}(t)\right].
       \label{eq:maxangle}
\end{equation}

 From the mathematical point of view, the calculation of the mean
times $t_{m}^{\sigma_{j}}(t;j)$ is a particular case of a general
problem, known in the theory of Markovian processes as the
first-passage time problem.\cite{Gardiner} In our case, the
Markovian process is the vector $\textbf{m}_{j}(t)$, and the level
set of first passages for $\textbf{m}_{j}(t)$ is the conical
surface defined by Eq.~(\ref{eq:maxangle}). We describe the
dynamics of the nanoparticle magnetic moments $\textbf{m}_{j}(t)
\equiv \textbf{m}_{j}$ by the system of stochastic Landau-Lifshitz
equations
\begin{equation}
       \dot\textbf{m}_{j}=-\gamma\textbf{m}_{j}\times
       (\textbf{H}_{j}+\textbf{n}_{j})
       -\frac{\lambda\gamma}{m}\,\textbf{m}_{j}\times(
       \textbf{m}_{j}\times\textbf{H}_{j}),
       \label{eq:LL}
\end{equation}
where $j=1,...,N$, $\gamma(>0)$ is the gyromagnetic ratio,
$\lambda(\ll1)$ is the damping parameter,
\begin{equation}
       \textbf{H}_{j}\equiv -\frac{\partial W_{j}(t)}{\partial
       \textbf{m}_{j}}=H_{a}[\cos\theta_{j}(t)+b_{j}(t)]
       \textbf{e}_{z}
       \label{eq:effield}
\end{equation}
is the effective magnetic field acting on $\textbf{m}_{j}$, and
$\textbf{n}_{j}=\textbf{n}_{j}(t)$ is the thermal magnetic field
that models the action of the thermostat. The thermal field is
approximated by Gaussian white noise with zero mean values
$\overline{\textbf{n}_{j}(t)}=0$ [the overbar denotes averaging
with respect to the sample paths of $\textbf{n}_{j}(t)$] and
correlations functions
\begin{equation}
       \overline{n_{i\alpha}(t_{1})n_{j\beta}(t_{2})}=
       2\Delta\delta_{ij}\delta_{\alpha\beta}\delta(t_{2}-t_{1}).
       \label{eq:corfunct}
\end{equation}
Here $n_{i\alpha}(t)$ ($\alpha=x,y,z$) are the Cartesian components
of $\textbf{n}_{i}(t)$, $\Delta=\lambda k_{B}T/\gamma m$ is the
intensity of the thermal magnetic field, $\delta_{ij}$ is the
Kronecker symbol, and $\delta(t)$ is the Dirac $\delta$ function.

If we treat the local dipolar fields $\textbf{h}_{j}(t)$ as
external magnetic fields, then we can consider the nanoparticles
to be independent. In other words, in this case the stochastic
Landau-Lifshitz equations (\ref{eq:LL}) are independent, and the
dynamics of each magnetic moment is described separately. Let
$P_{j}=P_{j}(\vartheta_{j},t|\vartheta_{j}',t')$ be the
conditional probability density that $\theta_{j}(t)=
\vartheta_{j}$ given that $\theta_{j}(t')= \vartheta_{j}'$ ($t\geq
t'$). [Note that in the case of axial symmetry $P_{j}$ does not
depend on the azimuthal angle of $\textbf{m}_{j}$.] Then, using
the Stratonovich interpretation\cite{Stratonovich} of
Eq.~(\ref{eq:LL}) and applying standard methods,\cite{Gardiner} we
can write for $P_{j}$ the forward Fokker-Planck equation
\begin{eqnarray}
       \frac{\partial P_{j}}{\partial t}&=&\frac{\partial}
       {\partial\vartheta_{j}}\bigg[\frac{\lambda\gamma}{m}
       \frac{\partial W_{j}(\vartheta_{j},t)}{\partial
       \vartheta_{j}}-\Delta\gamma^{2}\cot\vartheta_{j}
       \bigg]P_{j}\nonumber\\
       &&+\Delta\gamma^{2}\frac{\partial^{2}P_{j}}{\partial
       \vartheta_{j}^{2}}
       \label{eq:forwardFP}
\end{eqnarray}
and the backward Fokker-Planck equation
\begin{eqnarray}
       \frac{\partial P_{j}}{\partial t'}&=&\bigg[\frac{
       \lambda\gamma}{m}\frac{\partial W_{j}(\vartheta_{j}',t')}
       {\partial\vartheta_{j}'}-\Delta\gamma^{2}\cot\vartheta_{j}'
       \bigg]\frac{\partial P_{j}}{\partial\vartheta_{j}'}
       \nonumber\\
       &&-\Delta\gamma^{2}\frac{\partial^{2}P_{j}}{\partial
       \vartheta_{j}'^{2}}.
       \label{eq:backwardFP}
\end{eqnarray}

As a rule, the study of the magnetic properties of nanoparticle
ensembles is based on forward Fokker-Planck equations similar to
Eq.~(\ref{eq:forwardFP}), which allow us to express the
statistical characteristics of ensembles as functions of time $t$.
At the same time, backward Fokker-Planck equations are very useful
to describe the thermally induced reversal of the nanoparticle
magnetic moments.\cite{DY98} We use the backward Fokker-Planck
equation (\ref{eq:backwardFP}) to calculate the mean first-passage
times $t_{m}^{\sigma_{j}}(t;j)$.

To use Eq.~(\ref{eq:difference}) as the recurrence equation for
finding the reduced magnetization at the discrete times $t=t_{n}$
($n=0,1,...,M$, $t_{0}=0$, $t_{n+1}>t_{n}$), we need to calculate
$t_{m}^{\sigma_{j}} (t_{n};j)$ for $n=0,1,...,M-1$. Since to each
time $t_{n}$ corresponds the angle $\Omega_{j}(t_{n})$,  it is
necessary in Eq.~(\ref{eq:backwardFP}) to replace
$W_{j}(\vartheta_{j}',t')$ by $W_{j}(\vartheta_{j}', t_{n})$. In
other words, to find $t_{m}^{\sigma_{j}}(t_{n};j)$ we must use
Eq.~(\ref{eq:backwardFP}) with an energy term
$W_{j}(\vartheta_{j}',t')$ that $\textit{does not}$ depend on
$t'$. This important requirement results in a condition of
homogeneity for the random process $\theta_{j}(t)$,
$P_{j}(\vartheta_{j},t|\vartheta_{j}',t')= P_{j}(\vartheta_{j},
t-t'|\vartheta_{j}',0)$ and significantly simplifies the problem.

To calculate $t_{m}^{\sigma_{j}}(t_{n};j)$, we first introduce the
mean times $T_{j}=T_{j}^{\sigma_{j}}(\vartheta'_{j};t_{n})$, the time
necessary for $\theta_{j}(t)$ [$\theta_{j}(0)=\vartheta_{j}'$,
$\vartheta'_{j}\in(0,\Omega_{j}(t_{n}))$ if $\sigma_{j}=+$, and
$\vartheta'_{j}\in(\Omega_{j}(t_{n}),\pi)$ if $\sigma_{j}=-$]
to first reach the angle $\Omega_{j}(t_{n})$. The desired
times are expressed through $T_{j}$ as
\begin{equation}
       t_{m}^{\sigma_{j}}(t_{n};j)=T_{j}^{\sigma_{j}}(\pi(1-
       \sigma_{j}1)/2;t_{n}),
       \label{eq:represent}
\end{equation}
and the values $T_{j}$ themselves are represented in the form
\begin{eqnarray}
       T_{j}&=&\int_{0}^{\infty}du\int_{
       \Omega_{j}(t_{n})(1-\sigma_{j}1)/2}^{\pi(1-\sigma_{j}1)/2+
       \Omega_{j}(t_{n})(1+\sigma_{j}1)/2}d\vartheta
       \nonumber\\
       &&\times P_{j}(\vartheta,u|\vartheta'_{j},0).
       \label{eq:def}
\end{eqnarray}
Taking into account the initial condition $P_{j}(\vartheta_{j},
0|\vartheta_{j}',0)= \delta(\vartheta_{j}- \vartheta_{j}')$, the
homogeneity condition $P_{j}(\vartheta_{j},t|\vartheta_{j}',t')=
P_{j}(\vartheta_{j}, t-t'|\vartheta_{j}',0)$, and the expression
(\ref{eq:energy}), we obtain after integration of both sides of
the modified equation (\ref{eq:backwardFP}) over $u=t-t'$ and
$\vartheta= \vartheta_{j}$ as in Eq.~(\ref{eq:def})  the ordinary
differential equation for $T_{j}$
\begin{equation}
       \frac{d^{2}T_{j}}{d\vartheta_{j}'^{2}}+[\cot\vartheta_{j}'
       -2a(b_{j}(t_{n})+\cos\vartheta_{j}')\sin\vartheta_{j}']
       \frac{dT_{j}}{d\vartheta_{j}'}=-at_{r}
       \label{eq:eqT}
\end{equation}
($a=H_{a}m/2k_{B}T$, $t_{r}=2/\lambda\gamma H_{a}$).

To find the unique solution of Eq.~(\ref{eq:eqT}), we need to
impose two boundary conditions for the mean times $T_{j}$. The
first condition follows immediately from the definition of these
times: $T_{j}|_{\vartheta_{j}'=\Omega_{j}(t_{n})}=0$. We can find
the second by analysing the solutions of Eq.~(\ref{eq:eqT}) for
small vicinities of the angles $\vartheta_{j}'=0$ and
$\vartheta_{j}'=\pi$. There Eq.~(\ref{eq:eqT}) is reduced to
\begin{equation}
       \frac{d^{2}T_{j}}{d\vartheta_{j}'^{2}}+
       \frac{1}{\vartheta_{j}'-\pi(1-\sigma_{j}1)/2}\,
       \frac{dT_{j}}{d\vartheta_{j}'}=-at_{r},
       \label{eq:approxT}
\end{equation}
and its general solution is given by
\begin{eqnarray}
       T_{j}&=&c_{j}\ln{|\vartheta_{j}'-\pi(1-\sigma_{j}1)/2|}+d_{j}
       \nonumber\\
       &&-at_{r}[\vartheta_{j}'-\pi(1-\sigma_{j}1)/2]^{2}/4,
       \label{eq:approxsol}
\end{eqnarray}
where $c_{j}$ and $d_{j}$ are constants of integration. Since
$T_{j}$ are bounded quantities, the condition $c_{j}=0$ must hold.
This condition can be represented equivalently in the form of the
second boundary condition: $dT_{j}/d\vartheta_{j}'
|_{\vartheta_{j}'= \pi(1-\sigma_{j}1)/2}=0$. Note that these
boundary conditions correspond to the case where so-called
absorbing and reflecting barriers\cite{Gardiner} are placed at the
points $\vartheta_{j}'=\Omega_{j}(t_{n})$ and
$\vartheta_{j}'=0,\pi$, respectively.

Solving Eq.~(\ref{eq:eqT}) with these boundary conditions by the
method of variation of constants\cite{Arnold} and using the
representation (\ref{eq:represent}), we obtain the rigorous
formula
\begin{eqnarray}
       t_{m}^{\sigma_{j}}(t_{n};j)&=&at_{r}\int_{-\sigma_{j}b_{j}
       (t_{n})}^{1}dx\frac{e^{-a[x+\sigma_{j}b_{j}
       (t_{n})]^{2}}}{1-x^{2}}
       \nonumber\\
       &&\times \int_{x}^{1}dy\,e^{a[y+\sigma_{j}b_{j}(t_{n})]
       ^{2}},
       \label{eq:gentime}
\end{eqnarray}
which is valid for arbitrary $a$ and $|b_{j}(t_{n})|<1$. Using
Eq.~(\ref{eq:gentime}), let us calculate $w_{\sigma_{j}}(t_{n};j)$
for $\varepsilon_{j}\gg1$. According to Eq.~(\ref{eq:energy}), the
heights $\Delta U_{j}^{\sigma_{j}}(t)$ of the potential barrier
between the equilibrium directions of $\textbf{m}_{j}$ can be
written in the form $\Delta U_{j}^{\sigma_{j}}(t)=
\frac{1}{2}H_{a} m[1+\sigma_{j} b_{j}(t)]^2$, and, since $\Delta
U_{j}=\min{\Delta U_{j}^{\sigma_{j}}(t)}$, the condition
$\varepsilon_{j}\gg1$ leads to $a[1+\sigma_{j}
b_{j}(t_{n})]^2\gg1$. Taking into account that the asymptotic
formulas
\begin{eqnarray}
       \int_{x}^{1}dy\,e^{a[y+\sigma_{j}b_{j}(t_{n})]^{2}}
       &=&\frac{e^{a[1+\sigma_{j}b_{j}(t_{n})]^2}}
       {2a[1+\sigma_{j}b_{j}(t_{n})]},
       \nonumber\\
       \int_{-\sigma_{j}b_{j}(t_{n})}^{1}dx\frac{e^{-a[x+
       \sigma_{j}b_{j}(t_{n})]^{2}}}{1-x^{2}}&=&\frac{1}{2}
       \sqrt{\frac{\pi}{a}}\,\frac{1}{1-b_{j}^{2}(t_{n})}
       \nonumber
       \label{eq:asymp}
\end{eqnarray}
hold as $a[1+\sigma_{j} b_{j}(t_{n})]^2 \to \infty$, we find in
the same limit
\begin{eqnarray}
       w_{\sigma_{j}}(t_{n};j)&=&\frac{2}{t_{r}}\sqrt{
       \frac{a}{\pi}}\,[1-b_{j}^{2}(t_{n})][1+
       \sigma_{j}b_{j}(t_{n})]
       \nonumber\\
       &&\times e^{-a[1+\sigma_{j}b_{j}(t_{n})]^2}.
       \label{eq:probden}
\end{eqnarray}
Note that Eq.~(\ref{eq:probden}) follows also from Brown's
results\cite{B63} for isolated nanoparticles in a longitudinal
external field obtained with the forward Fokker-Planck equation
for $\varepsilon_{j}\gg1$. We have presented here an alternative
derivation of Eq.~(\ref{eq:probden}) based on the backward
Fokker-Planck equation, because within this approach the mean
first-passage times $t_{m}^{\sigma_{j}}(t;j)$ and the probability
densities of reorientation $w_{\sigma_{j}}(t_{n};j)$ are
calculated exactly for arbitrary $\varepsilon_{j}$.

If the condition $\varepsilon_{j}\gg1$ holds for all nanoparticles
and the ensemble state at time $t=t_{n}$  is known, then for the
same time we can find  the dipolar fields acting on each
nanoparticle, using the formula (\ref{eq:dipfield}), and calculate
the probability densities of reorientation of each magnetic
moment, using the formula (\ref{eq:probden}).

\subsection{Mean-field approximation}

To illustrate the influence of the correlations of the magnetic
moments and of the finite size of the nanoparticle ensemble on the
magnetic relaxation, we must first calculate the relaxation law
$\rho_{mf}(t)$ for an infinite lattice within the mean-field
approximation. To this end, we derive the equation that this
relaxation law satisfies, based on the results obtained above.
Since within the mean-field approximation the same mean dipolar
field acts on all magnetic moments, it is necessary in
Eq.~(\ref{eq:probden}) to replace $b_{j}(t)$ (we drop the index
$n$ in $t_{n}$) by $b(t)=\overline{b_{j}(t)}$. This implies that
all magnetic moments for which $\sigma_{j}=+$ and all magnetic
moments for which $\sigma_{j}=-$ are reoriented with the same
probability densities, $w_{+}(t)$ and $w_{-}(t)$ respectively,
where
\begin{equation}
       w_{\pm}(t)=\frac{2}{t_{r}}\sqrt{\frac{a}{\pi}}\,
       [1-b^{2}(t)][1\pm b(t)]\,e^{-a[1\pm b(t)]^2}.
       \label{eq:mf_probden}
\end{equation}
The function $b(t)$ is given by\cite{DT01}
\begin{equation}
       b(t)=-9.034\frac{m}{H_{a}d^{3}}\,\rho_{mf}(t),
       \label{eq:def_b}
\end{equation}
therefore the probability densities $w_{\pm}(t)$ depend on $t$
only via the reduced magnetization $\rho_{mf}(t)$, i.e.,
$w_{\pm}(t)= w_{\pm}(\rho_{mf}(t))$. Finally, using the equality
$\sum_{j}\sigma_{j} = N_{+}(t) - N_{-}(t)$ and the definition of
$\rho(t)$, we obtain from Eq.~(\ref{eq:difference}) for $\tau\to0$
and $N\to\infty$ the required differential equation
\begin{equation}
       \dot{\rho}_{mf}(t)=-\rho_{mf}(t)[w_{+}(t)+w_{-}(t)]-
       w_{+}(t)+w_{-}(t)
       \label{eq:mf(a)}
\end{equation}
($\rho_{mf}(0)=1$), which defines the law of magnetic relaxation
in the mean-field approximation. Note that the same equation
follows from the solution of the forward Fokker-Planck equation
(\ref{eq:forwardFP}).\cite{DT01}

Calling the right hand side of Eq. (\ref{eq:mf(a)})
$-F(\rho_{mf}(t))$, we can reduce this equation to the
integral form
\begin{equation}
       \int_{\rho_{mf}(t)}^{1}\frac{dx}{F(x)}=t.
       \label{eq:mf(b)}
\end{equation}
Its solution for small and large times yields\cite{DT01}
$\rho_{mf}(t)=1-t/\tau_{0}$ and
$\rho_{mf}(t)\propto\exp(-t/\tau_{\infty})$, respectively, where
\begin{equation}
       \tau_{0}=t_{r}\sqrt{\frac{\pi}{a}}\,\frac{e^
       {a(1-\xi)^2}}{4(1-\xi^2)(1-\xi)}
       \label{eq:tau_0}
\end{equation}
is the initial relaxation time,
\begin{equation}
       \tau_{\infty}=t_{r}\sqrt{\frac{\pi}{a}}\,\frac{e^{a}}
       {4[1+(2a-1)\xi]}
       \label{eq:tau_infty}
\end{equation}
is the final relaxation time, and $\xi=-b(0)$ ($0\leq\xi<1$) is a
parameter characterizing the intensity of dipolar interaction on
an infinite lattice. According to Eqs.~(\ref{eq:tau_0}) and
(\ref{eq:tau_infty}), the relaxation process in ensembles of
dipolar interacting nanoparticles is approximately characterized
by two relaxation times, $\tau_{0}$ and $\tau_{\infty}$, while in
the case of non-interacting nanoparticles, i.e., $\xi=0$, it is
characterized by the single relaxation time $\tau_{n}=
t_{r}\sqrt{\pi/16a}\exp a$. Since $\tau_{n}> \tau_{0}$ and
$\tau_{n}> \tau_{\infty}$, the dipolar interaction enhances
relaxation, and since $\tau_{0}< \tau_{\infty}$, the relaxation
rate decreases with time. For ensembles where the value of $\xi$
is not too small, the strong inequality $\tau_{0}\ll
\tau_{\infty}$ usually holds, and the decrease can be very large.

Note that the description of magnetic relaxation based on
Eqs.~(\ref{eq:difference}) and (\ref{eq:mf(a)}) is valid if the
quasiequilibrium distribution of the nanoparticle magnetic moments
is established, i.e., if\cite{DT01} $t\agt t_{qe}\sim at_{r}$. In
other words, these equations describe the slow phase of magnetic
relaxation. For $t\sim t_{qe}$, the probability of reorientation
of the nanoparticle magnetic moments from the initial state is
vanishingly small. Therefore we can transfer the origin of time to
an arbitrary point $t\sim t_{qe}$ and, since for $a>>1$ and $t\sim
t_{qe}$ the approximate equalities $\overline{m_{jz}(t)}\approx m$
hold, use the initial conditions $\rho(0)=1$ and $\rho_{mf}(0)=1$.

\section{Numerical Simulations}

\subsection{The computational algorithm}

According to the results of the previous section, to compute the
law of magnetic relaxation in some time interval $(0,t_{M})$ it is
necessary to know the states of the nanoparticle ensemble at the
discrete times $t=t_{n}$ ($n=0,1,...,M-1$). The state for $n=0$,
i.e., for $t=0$, is known from the initial condition:
$\sigma_{j}(0)=+$ for all $j$. To find the state at any other time
we proceed as follows. First we assume that the state of the
nanoparticle ensemble at the time $t=t_{n}$ is known. This means
that the set $A_{+}(t_{n})$ of numbers $j$ for which
$\sigma_{j}(t_{n})=+$, and the set $A_{-}(t_{n})$ of numbers $j$
for which $\sigma_{j}(t_{n})=-$ are fully defined. It is evident
that the set $A_{+}(t_{n})$ contains $N_{+}(t_{n})$ elements, and
the set $A_{-}(t_{n})$ contains $N_{-}(t_{n})$ elements.

Next, assuming that the time interval $\Delta t_{n+1}=
t_{n+1}-t_{n}$ is small enough, we introduce the average numbers
of reorientations
\begin{equation}
       \nu_{\pm}(t_{n},t_{n+1})=\Delta t_{n+1}\sum_{j\in A_{\pm}
       (t_{n})}w_{\pm}(t_{n};j),
       \label{eq:def_nu}
\end{equation}
that occur during $\Delta t_{n+1}$ for the sets of positively
(upper sign) and negatively (lower sign) oriented magnetic
moments. Strictly speaking, Eq.~(\ref{eq:def_nu}) is valid if the
strong inequality $\Delta t_{n+1}\max\{w_{\pm}(t_{n};j)\} \ll 1$
holds. Its use can drastically increase the time required for the
computation of the relaxation law in some cases. Therefore,
instead of the exact representation (\ref{eq:def_nu}) we use the
approximate one
\begin{equation}
       \nu_{\pm}(t_{n},t_{n+1})=\sum_{j\in A_{\pm}(t_{n})}
       U(\Delta t_{n+1}w_{\pm}(t_{n};j))
       \label{eq:approx_nu}
\end{equation}
[$U(x)=x$ if $x\leq1$, and $U(x)=1$ if $x>1$], which is valid if
the weaker condition $\nu_{\pm}(t_{n},t_{n+1}) \ll N$ holds, and
from Eq.~(\ref{eq:difference}) we obtain
\begin{equation}
       \rho(t_{n+1})=\rho(t_{n})-\frac{2}{N}[\nu_{+}(t_{n},t_{n+1})
       -\nu_{-}(t_{n},t_{n+1})].
       \label{eq:master_eq}
\end{equation}

Equations (\ref{eq:master_eq}), (\ref{eq:approx_nu}),
(\ref{eq:probden}), and (\ref{eq:dipfield}) allow us to calculate
the reduced magnetization at time $t=t_{n+1}$, if the nanoparticle
state at time $t=t_{n}$ is known. To find the nanoparticle state
at time $t=t_{n+1}$, we need to choose the sites where the
magnetic moments must be reoriented. To reflect the random
character of the thermal fluctuations, these sites should be
chosen randomly, while at the same time preference should be given
to those sites that have larger probabilities of reorientation. To
satisfy both requirements we proceed in the following way. First
we choose the time steps $\Delta t_{n+1}$. Since the number of
magnetic moments that are reoriented per unit time can appreciably
decrease with time, we select steps of varying length, $\Delta
t_{n+1}=\eta [w_{+}(t_{n}) + w_{-}(t_{n})] ^{-1}$. The parameter
$\eta$ must be chosen small enough to satisfy the condition
$\nu_{\pm}(t_{n},t_{n+1}) \ll N$ (in our calculations
$\eta=5\times10^{-3}$). Then we calculate the values $\Delta
t_{n+1} w_{+}(t_{n};j)$ for $j\in A_{+}(t_{n})$, and using the
formula (\ref{eq:approx_nu}) we find the average number of
reorientations
\begin{equation}
       \nu_{+}(t_{n},t_{n+1})=r_{+}(t_{n},t_{n+1})+
       \Delta t_{n+1}\sum_{j\in A_{+}'(t_{n})}w_{+}(t_{n};j),
       \label{eq:nu_plus}
\end{equation}
that occur during the time interval $\Delta t_{n+1}$ in the set of
positively oriented magnetic moments. Here $r_{+}(t_{n}, t_{n+1})$
is the number of lattice sites where $\Delta t_{n+1}
w_{+}(t_{n};j)>1$, and $A_{+}'(t_{n})$ is the set of lattice sites
where $\Delta t_{n+1} w_{+}(t_{n};j)\leq1$. Further, we introduce
the number of reorientations as $n_{+}(t_{n},t_{n+1})=
[\nu_{+}(t_{n}, t_{n+1})]+I$, where $[\nu_{+}(t_{n},t_{n+1})]$ is
the integer part of $\nu_{+}(t_{n},t_{n+1})$, and $I=0$ or 1 with
probability $p_{0}= \nu_{+}(t_{n},t_{n+1}) -[\nu_{+}(t_{n},
t_{n+1})]$ and $p_{1}=1-p_{0}$, respectively. Using a random
number generator, we  obtain a value for $n_{+}(t_{n},t_{n+1})$.

Among the $n_{+}(t_{n},t_{n+1})$ magnetic moments that must be
inverted at time $t=t_{n+1}$, we immediately invert the
$r_{+}(t_{n},t_{n+1})$ magnetic moments at lattice sites where the
condition $\Delta t_{n+1} w_{+}(t_{n};j)>1$ holds. [Recall that a
one-to-one correspondence exists between the lattice sites and
numbers $j$.] To find the remaining $n_{+}(t_{n},t_{n+1})
-r_{+}(t_{n},t_{n+1})$ lattice sites where the magnetic moments
have to be inverted, we first generate a random number that lies
in the interval of length $\sum_{j\in
A_{+}'(t_{n})}w_{+}(t_{n};j)$. This interval contains
$N_{+}(t_{n})- r_{+}(t_{n},t_{n+1})$ subintervals of lengths
$w_{+}(t_{n};j)$. We store the number $j$ of the subinterval
(i.e., the position of the site) that contains the random number
in memory, and then that subinterval is removed. Next we generate
a random number that lies in the new interval formed by the
remaining subintervals. The number $j$ of the subinterval that
contains this random number is again stored in memory, and then
this subinterval is also removed. Iterating this procedure
$n_{+}(t_{n}, t_{n+1}) - r_{+}(t_{n},t_{n+1})$ times, we find all
$n_{+}(t_{n},t_{n+1})$ lattice sites where positively oriented
magnetic moments must be inverted at time $t=t_{n+1}$.

Introducing in the same way the average number of reorientations
$\nu_{-}(t_{n}, t_{n+1})$ that occur in the set of negatively
oriented magnetic moments, and using the procedure described
above, we determine $n_{-}(t_{n}, t_{n+1})$ lattice sites where
these magnetic moments must be inverted at time $t=t_{n+1}$. Since
the ensemble state at time $t=t_{n}$ is known, the ensemble state
at $t=t_{n+1}$, i.e., after the inversion of $n_{+}(t_{n},
t_{n+1}) + n_{-}(t_{n}, t_{n+1})$ magnetic moments on well defined
lattice sites, is known too. Taking the latter state as the
initial state, we can find in the same manner the ensemble state
at time $t=t_{n+2}$, and so on.

Using the known state of the nanoparticle ensemble at the initial
time $t=0$ and applying the algorithm described above, we can find
the states for all times $t=t_{n}$ ($n=1,...,M-1$). Since our
algorithm is a probabilistic one, the reduced magnetization
calculated by the formula (\ref{eq:master_eq}) is a random
quantity. Let us designate that random reduced magnetization in
the $k$th numerical experiment as $\rho_{sim}^ {k}(t_{n})$. [A
numerical experiment consists of one application of the algorithm
to determine the ensemble states at all times $t=t_{n}$.] Then we
define the numerically simulated relaxation law as
\begin{equation}
       \rho_{sim}(t_{n})=\frac{1}{K}\sum_{k=1}^{K}
       \rho_{sim}^{k}(t_{n}),
       \label{eq:simul}
\end{equation}
where $K$ is the number of numerical experiments. To avoid any
misunderstanding, we emphasize that within the proposed algorithm
the dipolar field (\ref{eq:dipfield}) is calculated exactly, and
it is re-calculated after each time step.

\subsection{Numerical results and discussion}

We have used our analytical results and the numerical algorithm
described above to study the role that the finite size of the
nanoparticle ensemble and the correlations of the nanoparticle
magnetic moments play in magnetic relaxation. We found that the
reduced magnetization $\rho_{sim}(t)$ ($t>0$) decreases, when the
parameter $L$, a measure of the ensemble size, increases, i.e.,
$\rho_{sim}(t) |_{L_{1}} > \rho_{sim}(t) |_{L_{2}}$ if $L_{2} >
L_{1}$, and $\rho_{sim}(t) |_{L}$ tends to the limiting value
$\rho_{lim}(t)$ as $L\to\infty$. We explain such behavior of
$\rho_{sim}(t)$ as follows. Increasing $L$ leads to an increase,
on average, of the local dipolar fields acting on the nanoparticle
magnetic moments. As a result, the average of the probability
densities of reorientation of the positively oriented magnetic
moments increases, and the average of the probability densities of
reorientation of the negatively oriented magnetic moments
decreases. According to Eq. (\ref{eq:difference}), this means that
$\rho_{sim}(t)$ decreases when $L$ grows.

To verify this statement, we have calculated $\rho_{sim}(t)$ for
different ensembles of Co nanoparticles characterized by the
parameters $H_{a}=6400$\ Oe, $m/V=1400$\ G ($V$ is the
nanoparticle volume), $\lambda=0.2$, and $r=4$\ nm. As an
illustration, the function $\rho_{sim}(t)$, obtained at $T=300$\
K, $d=3r$, $L=50$, and $K=100$, and the approximate function
$\rho_{lim}(t)$ are shown in Fig.\ 2. We found the latter function
in the same way as $\rho_{sim}(t)$, but, to exclude boundary
effects, we assume that the basic nanoparticle ensemble (for which
we chose $L=100$) is surrounded by eight identical ensembles, and
each nanoparticle from the basic ensemble is considered as a
central one in the square box of the same size (i.e., $L=100$) and
interacts only with the nanoparticles which belong to this box. In
Fig.\ 2, we also show the function $\rho_{mf}(t)$ calculated via
the numerical solution of Eq.~(\ref{eq:mf(a)}) for an infinite
ensemble of Co nanoparticles with the same parameters. Note that
in this case $a\approx29.01$, $\xi\approx0.31$, $t_{r}\approx8.85
\times 10^{-11}$\ s, $\tau_{0}\approx 1.33\times 10^{-5}$\ s,
$\tau_{\infty}\approx 1.56$\ s, and $\tau_{n}\approx 28.89$\ s.

Since at $t=0$ the local dipolar field for an infinite ensemble is
always larger than the highest local dipolar field for a finite
one, the condition $\rho_{sim}(t)> \rho_{mf}(t)$ ($t>0$) must hold
for small enough times. We expect that the same condition holds
also for large enough times, since correlations of the
nanoparticle magnetic moments lead to slower magnetic relaxation
in the final phase than the mean-field theory predicts. As to the
relation between $\rho_{sim}(t)$ and $\rho_{mf}(t)$ at the
intermediate times, its character at a fixed temperature depends
on the ensemble size, i.e., on the parameter $L$.

To explain this dependence, we note first that at small times
magnetic relaxation for finite nanoparticle ensembles occurs
faster than in the case where the local dipolar fields are
replaced by their average value, i.e., the mean-field
approximation. Indeed, in the initial phase of magnetic relaxation
only a small number of the nanoparticle magnetic moments is
subjected to reorientation. In this case, the reoriented and most
of the non-reoriented magnetic moments are under the action of the
local dipolar fields, which exceed the mean dipolar field. This
means that $w_{-}(t;j) < w_{-}(t)$ for $j\in A_{-}(t)$,
$w_{+}(t;j) > w_{+}(t)$ for most  $j\in A_{+}(t)$, and therefore
the actual magnetic relaxation occurs faster than the mean-field
approximation predicts. [We emphasize that this conclusion is
valid for the initial phase of magnetic relaxation for finite as
well as infinite nanoparticle ensembles.] Furthermore, taking into
account that an increase in the size of the nanoparticle ensemble
leads to an increase, on average, of the local dipolar fields, we
expect the following behavior for the dependence of
$\rho_{sim}(t)$ on $L$ (for an illustration, see Fig.\ 2). If in
the nanoparticle ensemble the highest local dipolar field at $t=0$
is small enough in comparison to the case of an infinite ensemble,
i.e., if the parameter $L$ does not exceed the critical value
$L_{cr}=L_{cr}(T)$, then $\rho_{sim}(t) > \rho_{mf}(t)$ for all
$t>0$ (curve 1 in Fig.\ 2). At $L=L_{cr}$ the curves
$\rho_{sim}(t)$ and $\rho_{mf}(t)$ have a tangency point, and for
$L>L_{cr}$ they intersect at times $t=t_{1in}$ and $t=t_{2in}$
(curve 4, $t_{1in}\approx 2.25 \times 10^{-5}$\ s, $t_{2in}\approx
2.09 \times 10^{-2}$\ s). As $L$ is increased, the time $t_{1in}$
of the first intersection decreases, and the time $t_{2in}$ of the
second one increases. As a result, for $L\to\infty$ we have
$\rho_{sim}(t)\to \rho_{lim}(t)$, $t_{1in}\to 0$, and $t_{2in}$
tends to the limiting value $t_{in}$ (curve 2, $t_{in}\approx
0.46$\ s).

To characterize the difference between $\rho_{sim}(t)$ and
$\rho_{mf}(t)$, we introduce the parameter $\chi_{L}(t)=
[\rho_{sim}(t)-\rho_{mf}(t)]/\rho_{sim}(t)$. Its dependence on $t$
for the same ensembles of Co nanoparticles is shown in Fig.\ 3.
The nonzero value of $\chi_{L}(t)$ is caused by both the finite
size of the nanoparticle ensemble and the correlations of the
nanoparticle magnetic moments. Correlations significantly change
the relaxation law, and their role grows with time, i.e.,
$\chi_{L}(t) \to 1$ as $t\to\infty$.

The fact that the probability densities of reorientation
$w_{\sigma_{j}}(t;j)$, Eq.~(\ref{eq:probden}), depend
exponentially on the large parameter $a$ has two consequences. The
first is obvious, namely, the relaxation law $\rho_{sim}(t)$
strongly depends on temperature due to the inverse proportionality
of $a$ on $T$. The second is more complicated and refers to the
time dependence of $\rho_{sim}(t)$ and $\rho_{mf}(t)$ for
different $T$. According to the previous results, if at a certain
temperature the parameter $L$ satisfies the condition $L<L_{cr}$,
then $\rho_{sim}(t)> \rho_{mf}(t)$ for all $t>0$. As $T$
decreases, the probability densities $w_{\sigma_{j}}(t;j)$
decrease with different rates, and the smaller the temperature
becomes, the more their relative values differ. This means that as
$T$ is reduced, the reorientation of the nanoparticle magnetic
moments predominantly occurs at sites where $w_{\sigma_{j}}(t;j)$
are the largest. As a consequence, for small times the difference
between $\rho_{sim}(t)$ and the relaxation law derived by the
mean-field approximation grows as $T$ decreases. Therefore, if at
a given temperature the condition $L<L_{cr}$ holds and the values
of $L$ and $L_{cr}$ do not differ too much, then the curves
$\rho_{sim}(t)$ and $\rho_{mf}(t)$ can intersect at smaller
temperatures. The plots of $\rho_{sim}(t)$ calculated for
ensembles of Co nanoparticles for $L=50$ and $T=300$~K (see
Fig.~2), and for $L=50$ and $T=150$~K (see Fig.~4) demonstrate
this statement. In the latter case calculations yield $a\approx
58.02$, $\tau_{0}\approx 10.72$\ s, $\tau_{\infty} \approx 2.24
\times 10^{12}$\ s, $\tau_{n}\approx 8.11\times 10^{13}$\ s,
$t_{1in}\approx 56.12$\ s, and $t_{2in}\approx 6.83 \times
10^{11}$\ s.

The relaxation laws calculated above can not be determined using
the Monte Carlo method with time step quantification. According to
Ref.~\onlinecite{NCK00}, the time interval $\Delta t$ that
corresponds to one Monte Carlo step is written in our notations as
\begin{equation}
       \Delta t=\frac{R^{2}(1+\lambda^{2})m}{20k_{B}T\lambda
       \gamma}
       \label{eq:timestep}
\end{equation}
($R<1$), and  the number $M=\mu\tau_{n}/\Delta t$ of the Monte
Carlo steps that are necessary to calculate the relaxation law on
the time interval $(0,\mu\tau_{n})$ is given by
\begin{equation}
       M=\frac{5\mu}{R^{2}(1+\lambda^{2})}\sqrt{\frac{\pi}{a^{3}}}\,
       e^{a}.
       \label{eq:steps}
\end{equation}
For the nanoparticle ensembles considered here,
Eq.~(\ref{eq:steps}) for $R=1$ and $\mu=0.2$ yields
$M\approx4.33\times10^{10}$ for $T=300$\ K, and
$M\approx6.08\times10^{22}$ for $T=150$\ K. Such values of $M$
render of course the use of that method impractical. For
comparison, in our approach the number $M$ of time steps $\Delta
t_{n+1}$, defined by the condition $\sum_{m=1}^{M}\Delta
t_{m}=\mu\tau_{n}$, equals 157 and 169, respectively.

\section{Conclusions}

We have developed a new method for the numerical simulation of
thermally activated magnetic relaxation in  2D ensembles of
uniaxial ferromagnetic nanoparticles whose easy axes of
magnetization are perpendicular to their distribution plane. It is
based on the analytical determination of the probability densities
of reorientation of the nanoparticle magnetic moments and on the
numerical determination of the nanoparticle ensemble states for a
discrete sequence of times. Using the backward Fokker-Planck
equation, we have formulated a rigorous approach to calculate
those probability densities, and in the case of high potential
barriers between the equilibrium directions of the nanoparticle
magnetic moments we have studied the law of magnetic relaxation by
this method.

We have shown that magnetic relaxation in finite nanoparticle
ensembles can differ strongly from that predicted by the
mean-field approximation for infinite ensembles. This difference
is caused by the finiteness of the ensemble size as well as
correlations between the magnetic moments, which result from the
dipolar interaction between nanoparticles. In a finite ensemble,
magnetic relaxation for small and large times occurs slower than
the mean-field theory predicts for infinite ensembles, and for
intermediate times the corresponding relaxation curves, depending
on the ensemble size and temperature, can intersect twice.
Increase of the ensemble size enhances relaxation, and in the
limiting case of an infinite ensemble, magnetic relaxation for
small times occurs faster and for large times slower than for the
mean-field theory. This feature of the relaxation law is caused by
the correlation effects whose role grows with time.

\section*{ACKNOWLEDGMENTS}

We are grateful to Werner Horsthemke for the critical reading of
this manuscript and his valuable comments. This work was supported
in part by NATO Grant No. PST.CLG.978108.

\cleardoublepage

\begin{figure}[htbp]
        \centering
        \includegraphics{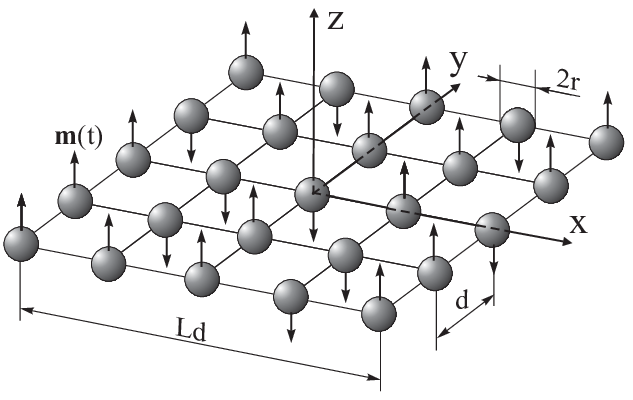}
        \caption{\label{fig1}Schematic representation of the
         2D nanoparticle ensemble.}
\end{figure}

\begin{figure}[htbp]
        \centering
        \includegraphics[width=3in,height=3in]{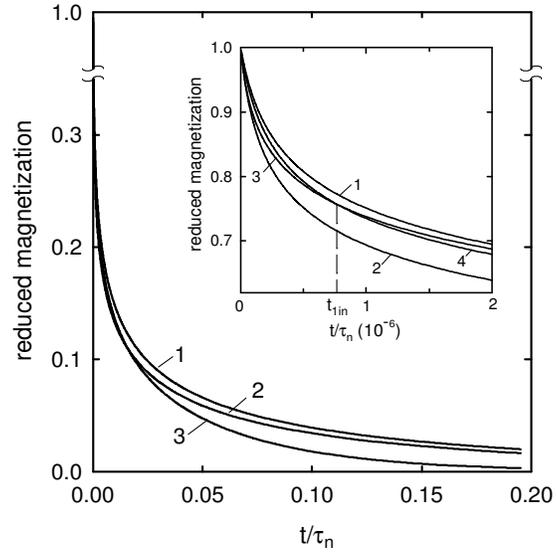}
        \caption{\label{fig2}Plots of $\rho_{sim}(t)$ for $L=50$ (curve
        1), $\rho_{lim}(t)$ (curve 2), and $\rho_{mf}(t)$ (curve 3).
        Inset: The same plots and the plot of $\rho_{sim}(t)$ for
        $L=70$ (curve 4) for small times.}
\end{figure}

\begin{figure}[htbp]
        \centering
        \includegraphics[width=3in,height=3in]{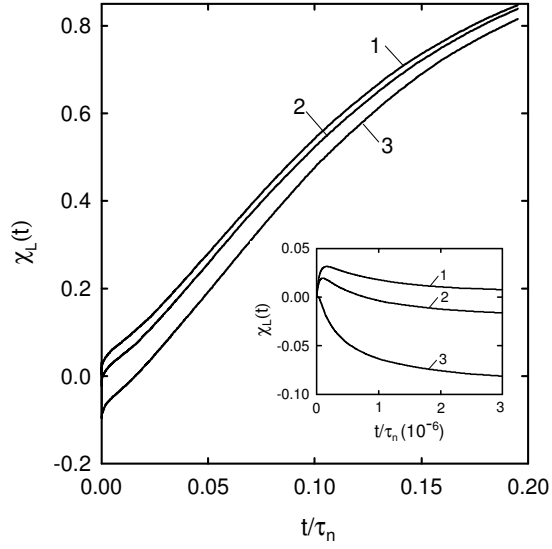}
        \caption{\label{fig3}Plots of $\chi_{L}(t)$ for $L=50$ (curve 1),
        $L=70$ (curve 2), and $L=\infty$ (curve 3). Inset: The same
        plots for small times.}
\end{figure}

\begin{figure}[htbp]
        \centering
        \includegraphics[width=3in,height=3in]{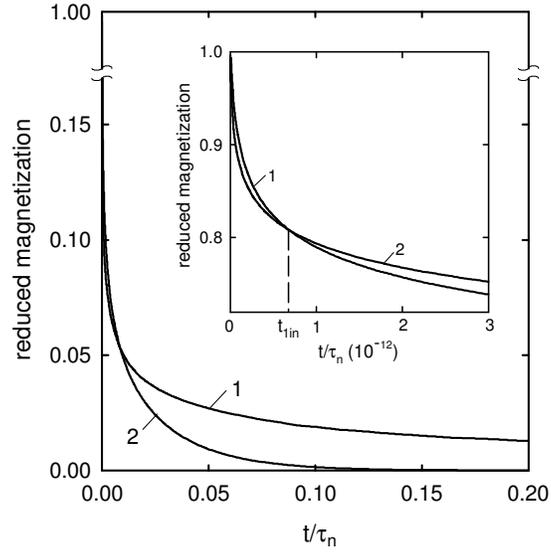}
        \caption{\label{fig4}Plots of $\rho_{sim}(t)$ (curve 1) and
        $\rho_{mf}(t)$ (curve 2) for $L=50$ and $T=150$\ K.
        Inset: The same plots for small times.}
\end{figure}
\end{document}